\documentclass{eptcs}
\providecommand{\event}{TFPIE 2022}

\usepackage{fontaxes}
\usepackage{graphicx}
\usepackage{stfloats}
\usepackage{multido}
\usepackage{setspace}
\usepackage{textcomp}
\usepackage{alltt}
\usepackage{textcomp}
\usepackage{pgfplots}
\pgfplotsset{compat=newest}
\usepackage{pgfplotstable}
\usepackage{cleveref}
\usepackage{doi}

\newcommand{\htdp}{\texttt{HtDP}}
\newcommand{\racket}{\texttt{Racket}}
\newcommand{\oop}{\texttt{OOP}}
\newcommand{\arrow}{\(\rightarrow\)}
\newcommand{\lexp}{$\lambda$-expression}
\newcommand{\lexps}{$\lambda$-expressions}
\newcommand{\elist}{\textquotesingle{()}}

\title{Introduction to Functional Classes in CS1}

\author{Marco T. Moraz\'an
\institute{Seton Hall University}
\email{morazanm@shu.edu}}
\def\titlerunning{Introduction to Functional Classes in CS1}
\def\authorrunning{M. T. Moraz\'an}

\begin{document}

\maketitle

\begin{abstract}
Students introduced to programming using a design-based approach and a functional programming language become familiar with first-class functions. They rarely, however, connect first-class functions to objects and object-oriented program design. This is a missed opportunity because students inevitably go on to courses using an object-oriented programming language. This article describes how students are introduced to objects within the setting of a design-based introduction to programming that uses a functional language. The methodology exposes students to interfaces, classes, objects, and polymorphic dispatch. Initial student feedback suggests that students benefit from the approach.
\end{abstract}

\section{Introduction}
Design-based introduction to programming courses using a functional programming language, as put forth in the textbook \emph{How to Design Programs} (\htdp), are well-established and have become popular. In such courses students learn type-based design through the use of \emph{design recipes}--a series of steps, each with a concrete result, that take the student from a problem statement to a well-designed and tested solution. Among other things, students are exposed to first-class functions. That is, students learn that functions may be data (i.e., functions may be passed to and returned by a function). For instance, the following \racket \ function returns a function for the composition of, \texttt{f} and \texttt{g}, two given functions:
\begin{alltt}
     ;; (X \(\rightarrow\) Y) (Y \(\rightarrow\) Z) \(\rightarrow\) (X \(\rightarrow\) Z)
     ;; Purpose: Return a function for (f \(\circ\) g)(x)
     (define (compose-f-g f g)
       ;; X \(\rightarrow\) Z
       ;; Purpose: Return (f \(\circ\) g)(x) for the given X-value
       (\(\lambda\) (an-x) (f (g an-x))))

     ;; number \(\rightarrow\) number
     ;; Purpose: Add 2 to the given number
     (define add2 (compose-f-g add1 add1))

     ;; number \(\rightarrow\) number
     ;; Purpose: Return the given number
     (define id   (compose-f-g add1 sub1))

     (check-expect (add2 3)  5)
     (check-expect (id 10)  10)
\end{alltt}
The signature states that the first input is a function that maps an element of type \texttt{X} to an element of type \texttt{Y}, that the second input is a function that maps an element of type \texttt{Y} to an element of type \texttt{Z}, and that the returned value is a function that maps an element of type \texttt{X} to an element of type \texttt{Z}. The function header has two parameters \texttt{f} and \texttt{g} in accordance with the signature. The body is a $\lambda$-expression that defines a function that takes as input a value of type \texttt{X} and returns a value of type \texttt{Z}. This $\lambda$-expression implements \texttt{(f $\circ$ g)(x)} as stated in its purpose statement. It has a parameter, \texttt{an-x}\footnote{Students are encouraged to use variable names that help readers of the code understand what is being represented. In this case, \texttt{an-x} suggests a value of type \texttt{X}.}, and applies \texttt{f} to the result of applying \texttt{g} to \texttt{x}. Observe that the returned function satisfies \texttt{compose-f-g}'s signature. Tests are written for functions created using \texttt{compose-f-g}. In the example above, two functions are defined: one to add 2 to its input and the identity function for numbers. Test are written for these functions using \texttt{check-expect}\footnote{The forms \texttt{check-expect} and \texttt{check-within}, among others, are used by students to write unit tests in the \racket \ student languages.}.

The other side of the coin is rarely explored to any depth. That is, thinking of data as a function is not emphasized. This is a missed opportunity to introduce beginners to object-oriented concepts that they will invariably see in future courses. Such an introduction is important because it facilitates the transition to object-oriented programming (\oop) and design. Furthermore, it helps address the concerns of \oop \ instructors that worry about beginners not being exposed to object-oriented programming in their introduction courses. At the heart of the idea is having a returned function implement an interface and use message-passing to provide services. Students are started with the implementation of compound data of finite size that does not have variety (i.e., no subtypes) and then are moved onto implementing union types (i.e., with subtypes) including compound data of arbitrary size.

The article is organized as follows. \Cref{rw} briefly discusses related work. \Cref{sb} discusses the students' background. \Cref{is} discusses how students are introduced to interfaces, classes, and objects by implementing structures using functions. \Cref{dr} presents a design recipe for interfaces. \Cref{iu} discusses the implementation of union types and polymorphic dispatch using the design recipe for interfaces. \Cref{fb} presents initial student feedback. Finally, \Cref{fw} presents conclusions and directions for future work.

\section{Related Work}
\label{rw}

The most common source students have to introduce them to objects are \oop \ textbooks. Such textbooks traditionally emphasize that classes and objects combine code and data to be manipulated as illustrated by the following examples:
\begin{itemize}
  \item An object is a structure for incorporating data and the procedures for working with that data \cite{clark}.
  \item A class is a group of objects that share common state and behavior. A class is an abstraction or description of an object. An object, on the other hand, is a concrete entity that exists in space and time \cite{tymann}.
  \item Every object is an instance of a class, which serves as the type of the object and as a blueprint, defining the data which the object stores and the methods for accessing and modifying that data \cite{goodrich}.
  \item Objects are used to combine data with procedures that operate on that data \cite{preiss}.
\end{itemize}
As the reader can appreciate \oop \ textbooks emphasize the encapsulation of data and methods to define classes and objects. These textbooks then proceed with examples illustrating how to use classes and how to create objects. The unstated assumption is that beginning students absorb design principles by studying code samples. Like the approach taken by many \oop \ textbooks the work presented in this article emphasizes the definition of behavior. Students are taught to combine data and functions using encapsulation to create a package called a class. In contrast, however, program design and the proper use of \oop \ abstractions are emphasized from the beginning. That is, behavior is defined in an interface, an implementation of an interface is called a class, and an instance of a class is an object. Observe that such an approach truly separates specification from implementation in the mind of beginners.

The unpublished textbook \texttt{How to Design Classes} (\texttt{HtDC}) \cite{HtDC} is closely coupled with a domain specific language (\texttt{DSL}) known as \texttt{ProfessorJ} \cite{ProfJ}. It first introduces students to classes as a mechanism to define a compound datatype that may have many fields (similar to a structure). Students are explained that every class has a constructor that initializes the fields and produces an object--an instance of the type defined by the class. The design of a class starts by understanding how data is to be represented. Union types are introduced to motivate the need for interfaces. In essence, an interface is used to define a type and glue together its subtypes. The design of methods is not tackled until students develop some expertise defining classes that only have a single constructor. Similar to \texttt{HtDC}, the approach described in this article starts with parallels between structures and classes with no subtypes and then moves to union types. In contrast, however, we use \racket \ syntax (specifically, the intermediate student language with \texttt{lambda}) and rely quite heavily on the familiarity students have developed with this syntax to introduce them to \oop \ concepts. That is, a new specialized \texttt{DSL} is not required.

Another design-based approach developed at Northeastern University teaches students (that have used \htdp \ in their first course) using a hybrid approach \cite{PwC}. The beginning of the second course proposes the use of several \texttt{DSL}s embedded in \racket \ before making the full transition into \texttt{Java}. The motivation for such an approach is that students first develop a command of basic \oop \ principles before tackling \texttt{Java}'s syntax, types, and compiler (as opposed to an interpreter). This approach also relies heavily of the design recipe to provide students (and instructors) with a framework for program design and discussion. From the start objects are described as consisting of data and functions that define behavior. A class is first thought of as a structure and its fields are accessed using a dot notation. For example, \texttt{(point . x)} denotes that the message, \texttt{x}, on the right hand of the dot is used to request a value from the object, \texttt{point}, on the left hand side of the dot. Union types and design based on structural recursion are introduced as requiring a distinct class for each subtype. The transition to \texttt{Java} occurs in the middle of the semester. This transition is made to functional \texttt{Java} first to make programs as similar as possible to those developed using the \oop \ \texttt{DSL}s embedded in \racket. Later mutation and looping constructs are presented. In a similar spirit, the work presented in this article introduces objects as encapsulated data and functions. In addition, it also starts with parallelism between structures and classes and then moves to union types. In contrast, the work presented here is not part of an \oop \ course. Instead, the work presented is intended to introduce students to object-oriented abstractions within the context of a first programming course using the program by design methodology put forth in \htdp.

\section{Student Background}
\label{sb}

At Seton Hall University, the introductory Computer Science 2-course sequence focuses on problem solving using a computer \cite{mtm22,mtm25}. The languages of instruction are the successively richer subsets of \racket \ known as the student languages which are tightly-coupled with \htdp \ \cite{HtDP,HtDP2}. No prior experience with programming is assumed. The first course starts by familiarizing students with primitive data, primitive functions, and library functions to manipulate images (i.e., the image teachpack). During this introduction, students are taught about variables, defining their own functions, and the importance of writing signatures, purpose statements, and tests. The next step of the course introduces students to data analysis and programming with compound data of finite size (i.e., structures). At this point, students are introduced to the first design recipe. Building on this experience, students develop expertise in processing compound data of arbitrary size such as lists, natural numbers, and trees. In this part of the course, students learn to design functions using structural recursion. After structural recursion, students are introduced to functional abstraction and the use of higher-order functions such as \textsf{map} and \textsf{filter}. The first semester ends with a module on distributed programming \cite{mtm26,mtm27}.

As part of the module on functional abstraction students are introduced to \lexps \ and curried functions. Both of these topics are pivotal to introducing students to interfaces and objects. Students understand that a \lexp \ evaluates to a function that consumes input and returns a value. With this understanding students are introduced to curried functions and how they provide the convenience of not consuming all of their input at once. For example, consider a set of functions as the following to scale a list of numbers:
\begin{alltt}
     (define (scale-by-2 L)
       (map (\texttt{\(\lambda\)} (x) (* 2 x)) L))

     (define (scale-by-5 L)
       (map (\texttt{\(\lambda\)} (x) (* 5 x)) L))

     (define (scale-by-3 L)
       (map (\texttt{\(\lambda\)} (x) (* 3 x)) L))
            \(\vdots\)
\end{alltt}
Observe that there is a great deal of repetition among the functions. Students learn to eliminate such repetitions using functional abstraction. For the set of functions above, the result of functional abstraction is a curried function that takes as input the scaling factor and returns a function to scale a list by that factor as follows:
\begin{alltt}
   ;; number \arrow ((listof number) \arrow (listof number))
   (define (make-list-scaler sc)
     (\(\lambda\) (L) (map (\(\lambda\) (x) (* sc x)) L)))
\end{alltt}
The original definitions may be refactored and tested as follows:
\begin{alltt}
     (define scale-by-2 (make-list-scaler  2))

     (define scale-by-5 (make-list-scaler  5))

     (define scale-by-3 (make-list-scaler  3))
        \(\vdots\)

     (check-expect (scale-by-2 \elist) \elist)
     (check-expect (scale-by-2 \textquotesingle{(1 2 3)}) \textquotesingle{(2 4 6)})
     (check-expect (scale-by-5 \elist) \elist)
     (check-expect (scale-by-5 \textquotesingle{(4 0)}) \textquotesingle{(20 0)})
     (check-expect (scale-by-3 \elist) \elist)
     (check-expect (scale-by-3 \textquotesingle{(-2 8)}) \textquotesingle{(-6 24)})
        \(\vdots\)
\end{alltt}
Most functional programmers are likely to consider \texttt{make-list-scaler} a function that returns a function. This is technically correct but it can also be described using \oop \ lingo. We may also say that \texttt{make-list-scaler} is a \emph{class} that returns \texttt{objects} that scale a list of numbers by a given scalar. This connection is not automatically made by students in the course. It is, therefore, necessary to make this connection explicitly in class. The result is twofold: students gain more interest in the techniques being taught given that they see how they apply beyond their first course and students arrive at their first \oop \ course with an understanding of some of the abstractions object-oriented languages emphasize.

\section{Structures as Interfaces}
\label{is}

\begin{figure*}[t!]
\begin{alltt}
     ;; A 3Dposn is a structure: (make-3Dposn number number number)
     (define-struct 3Dposn (x y z))

     ;; Sample instances of 3Dposn
     (define ORIGIN  (make-3Dposn 0 0 0))
     (define A3DPOSN (make-3Dposn 2 3 5))

     ;; 3Dposn \arrow number
     ;; Purpose: Return the distance to the origin of the given 3Dposn
     (define (dist-origin a-3dposn)
       (sqrt (+ (sqr (3Dposn-x a-3dposn))
                (sqr (3Dposn-y a-3dposn))
                (sqr (3Dposn-z a-3dposn)))))

     ;; Sample expressions for dist-origin
     (define ORIGIND  (sqrt (+ (sqr (3Dposn-x ORIGIN))
                               (sqr (3Dposn-y ORIGIN))
                               (sqr (3Dposn-z ORIGIN)))))
     (define A3DPOSND (sqrt (+ (sqr (3Dposn-x A3DPOSN))
                               (sqr (3Dposn-y A3DPOSN))
                               (sqr (3Dposn-z A3DPOSN)))))

     ;; Tests for dist-origin
     (check-within (dist-origin ORIGIN)  ORIGIND  0.01)
     (check-within (dist-origin A3DPOSN) A3DPOSND 0.01)
     (check-within (dist-origin (make-3Dposn 10 20 30))  37.42 0.01)
\end{alltt}
\caption{Distance to the Origin Program for a \texttt{3Dposn}.} \label{3dposn1}
\end{figure*}

Students are reminded that a data definition defines a type and in the case of compound data, like a structure, it also defines the types of the components. It states nothing about the valid type operations. The valid operations on an instance of the type may be defined in an \emph{interface}. An interface defines the behavior of a defined type. In other words, an interface specifies the operations that are valid on a type. Students are asked to consider, for example, the problem of computing the distance to the origin for a given point on a three-dimensional plane. Following the steps of the design recipe yields the program displayed in \Cref{3dposn1}. There are a few unstated assumptions in the program. First, it is assumed that there is a function to construct a \texttt{3Dposn}. Second, it is assumed that there are selector functions for the components of a \texttt{3Dposn}. Third, the values of \texttt{x}, \texttt{y}, \texttt{z}, and the constructor/selector functions are stored separately from the function \texttt{dist-origin}. The fact that students know that constructor and selector functions exists is only because they have been told they are created for them when they define a structure.

In addition to defining a type, an interface is developed to explicitly define the expected behavior. An interface outlines the valid operations and the returned type. For a \texttt{3Dposn} the interface is:
\begin{description}
  \item[\ \ \ \ \ Request \textbf{\texttt{x}}:] \texttt{number}
  \item[\ \ \ \ \ Request \textbf{\texttt{y}}:] \texttt{number}
  \item[\ \ \ \ \ Request \textbf{\texttt{z}}:] \texttt{number}
  \item[\ \ \ \ \ Request \textbf{\texttt{distance}}:]  \texttt{number}
\end{description}
The interface makes it clear to any reader or user which are the valid operations on a \texttt{3Dposn}. Observe that an interface says nothing about how a data type and its valid operations are implemented.

Students are asked to take a moment to ponder what has just been done. The data definition and the interface together explicitly relate \texttt{x}, \texttt{y}, \texttt{z}, \texttt{3Dposn-x}, \texttt{3Dposn-y}, \texttt{3Dposn-z}, and \texttt{dist-origin}. If they are related then we ought to be able to encapsulate them into a single package. Whenever a \texttt{3Dposn} is constructed the package returned ought to be able to perform all the operations in the interface. Encapsulation is a practice that students are familiar with given that they have previously been exposed to \texttt{local}-expressions.

Students are explained that to provide functionality a technique called \emph{message-passing} is used. An interface is implemented by a constructor function called a \emph{class} that returns a message-processing function. This is a curried function that receives as input a message requesting a service. For example, this function may get the message \texttt{\textquotesingle{getx}} requesting the \texttt{x} value of the \texttt{3Dposn}. An interface, therefore, must specify the messages used to request a service. We can now refine the \texttt{3Dposn} interface to be:
\begin{alltt}
     \textquotesingle{getx}: number
     \textquotesingle{gety}: number
     \textquotesingle{getz}: number
      \textquotesingle{d2o}: number
\end{alltt}
Observe that there is a unique message (in this case a symbol) associated with each service. The idea is that the message-processing function determines what value to compute by examining the message it gets as input. Note that embedded in the interface definition is a data definition for a \texttt{message}. A message is an enumeration type: either \texttt{\textquotesingle{getx}}, \texttt{\textquotesingle{gety}}, \texttt{\textquotesingle{getz}}, or \texttt{\textquotesingle{d2o}}.

\begin{figure*}[t!]
\begin{alltt}
     (require 2htdp/abstraction)

     ;; number number number \arrow 3Dposn
     ;; Purpose: Return a \texttt{3Dposn} object
     (define (make-3Dposn x y z)
       (local [;; 3Dposn \arrow number Purpose: Return distance to origin
               (define (dist-origin x y z) (sqrt (+ (sqr x) (sqr y) (sqr z))))
               ;; message \arrow 3Dposn service throws error
               ;; Purpose: To manage messages for a 3Dposn
               (define (manager m)
                 (match m
                   [\textquotesingle{getx} x]
                   [\textquotesingle{gety} y]
                   [\textquotesingle{getz} z]
                   [\textquotesingle{d2o}  (dist-origin x y z)]
                   [else (error (string-append "Unknown message to 3Dposn: "
                                               (symbol->string m)))]))]
         manager))

     ;; Sample 3Dposn objects
     (define ORIGIN  (make-3Dposn 0 0 0))
     (define A3DPOSN (make-3Dposn 2 3 5))

     ;; Tests for 3Dposn
     (check-within (ORIGIN  \textquotesingle{getx}) 0    0.01)
     (check-within (A3DPOSN \textquotesingle{gety}) 3    0.01)
     (check-within (ORIGIN  \textquotesingle{getz}) 0    0.01)
     (check-within (A3DPOSN \textquotesingle{d2o})  6.16 0.01)
     (check-error (A3DPOSN \textquotesingle{move-r})
                  "Unknown message to 3Dposn: move-r")
\end{alltt}
\caption{Interface Implementation for \texttt{3Dposn}.} \label{3dposn2}
\end{figure*}

Students are explained that a constructor function (i.e., class) that implements an interface encapsulates the values of and the operations on a type. It defines a constructor for instances of a type. The value returned by a class is called an \emph{object}. An object is an instance of an interface and knows how to perform all the services in the interface using message-passing. The \texttt{3Dposn} class is displayed in \Cref{3dposn2}. The class takes as input 3 numbers and returns a \texttt{3Dposn}. It is named \texttt{make-3Dposn} to easily identify its role as a constructor for \texttt{3Dposn}s. Its body is a \texttt{local}-expression that defines an auxiliary function for any value that needs to be computed (in this case only distance to origin) and the message-processing function called \texttt{manager}. The \texttt{manager} takes as input a message and returns (the value of) a service defined in the interface. The body of \texttt{manager} is a \texttt{match}-expression to distinguish the message varieties. If a service requires no computation a value is directly returned. If computation is required (like computing the distance to the origin) a local function is called. In this case, \texttt{manager} is a guarded function that throws an error when the received input is not a \texttt{message}.  We can observe that \texttt{manager} is also an object given that it knows how to compute all the services in the interface and, therefore, the \texttt{local}-expression returns it.

Testing interfaces requires defining one or more objects and writing tests to check that services are correctly provided. In \Cref{3dposn2} two \texttt{3Dposn} objects are defined. The tests check the result obtained from passing each message to an object. For example, the \texttt{x}-coordinate of \texttt{ORIGIN} is obtained using \texttt{(ORIGIN  \textquotesingle{getx})}--passing the message  \texttt{\textquotesingle{getx}} to \texttt{ORIGIN} (akin to the \texttt{Java} syntax: \texttt{ORIGIN.getx()}). The expected value is 0.

Students suddenly realize that a \texttt{3Dposn}, which they have always thought of as data, is a function. Specifically, it is an instance of the curried function \texttt{manager} (i.e., an object) that is specialized for the values of \texttt{x}, \texttt{y}, \texttt{z} given to \texttt{make-3Dposn}. \texttt{ORIGIN} is a \texttt{3Dposn} object in which \texttt{x} = \texttt{y} = \texttt{z} = 0. \texttt{A3DPOSN} is a \texttt{3Dposn} object in which \texttt{x} = 2, \texttt{y} = 3, and \texttt{z} = 5. Students clearly see that, just like functions can be data, data can be a function and an object \emph{is} a function.

\subsection{Improving the Human Interface}

\begin{figure*}[t!]
\begin{alltt}
     ;; 3Dposn \arrow number
     ;; Purpose: Return the x of the given 3Dposn
     (define (3Dposn-x a-3dposn) (a-3dposn \textquotesingle{getx}))

     ;; Tests for 3Dposn-x
     (check-within (3Dposn-x ORIGIN)  0  0.01)
     (check-within (3Dposn-x A3DPOSN) 2 0.01)
     (check-within (3Dposn-x (make-3Dposn 10 20 30)) 10 0.01)

     ;; 3Dposn \arrow number
     ;; Purpose: Return the y of the given 3Dposn
     (define (3Dposn-y a-3dposn) (a-3dposn \textquotesingle{gety}))

     ;; Tests for 3Dposn-y
     (check-within (3Dposn-y ORIGIN)  0  0.01)
     (check-within (3Dposn-y A3DPOSN) 3  0.01)
     (check-within (3Dposn-y (make-3Dposn 10 20 30)) 20 0.01)

     ;; 3Dposn \arrow number
     ;; Purpose: Return the z of the given 3Dposn
     (define (3Dposn-z a-3dposn) (a-3dposn \textquotesingle{getz}))

     ;; Tests for 3Dposn-z
     (check-within (3Dposn-z ORIGIN)  0  0.01)
     (check-within (3Dposn-z A3DPOSN) 5 0.01)
     (check-within (3Dposn-z (make-3Dposn 10 20 30)) 30 0.01)

     ;; 3Dposn \arrow number
     ;; Purpose: Return distance to origin of given 3Dposn
     (define (dist-origin a-3dposn) (a-3dposn \textquotesingle{d2o}))

     ;; Tests for dist-origin
     (check-within (dist-origin ORIGIN)  0    0.01)
     (check-within (dist-origin A3DPOSN) 6.16 0.01)
     (check-within (dist-origin (make-3Dposn 10 20 30)) 37.42 0.01)
\end{alltt}
\caption{Wrapper Functions for \texttt{3Dposn}.} \label{3dposn-wrappers}
\end{figure*}

Message-passing may reduce the readability of the code. For example, does \texttt{(A3DPOSN \textquotesingle{d20})} communicate to others that this expression represents \texttt{A3DPOSN}'s distance to the origin? Unless you are intimately familiar with the message-passing protocol it is likely that this expression is meaningless. Furthermore, it is unlikely that any programmer (including the student that wrote it) will permanently remember the message-passing protocol in the near and far future. This will make it unnecessarily more difficult to refine the program.

To mitigate this problem, \emph{wrapper functions} for the services provided by an interface may be written. A wrapper function hides the details of the implementation. In this case, it hides the details of message-passing. The idea is to allow programmers to use \texttt{3Dposn}s without forcing them to know how they are implemented. A wrapper function is needed for each service in the interface. It takes as input an object (and any additional inputs if any) and its body applies the object to the appropriate message. Wrapper functions are designed following the steps of the design recipe.

\Cref{3dposn-wrappers} displays the wrapper functions for \texttt{3Dposn}. Adding this code to the one displayed in \Cref{3dposn2} allows programmers to use a nicer version of the defined interface. Instead of explicitly using message-passing, they can use the wrapper functions. Observe that now programmers have the same interface as the one used in \Cref{3dposn1}. Writing wrapper functions does not provide a programmer with new computational powers, but it is an abstraction that liberates a programmer from the details of a message-passing protocol.

\subsection{Services that Require More Input}

After an interface is implemented it may be necessary to add more services. This means expanding the message-processing function and, if necessary, designing (local) auxiliary functions. If a value may be computed using only the information stored in an object then adding a service is the same as what is done for \texttt{dist-origin} for \texttt{3Dposn}. For example, adding a service to determine if a given \texttt{3Dposn} object is on the \texttt{x}-axis may de done by comparing the \texttt{x}-value to 0.

If a service requires further input the answer cannot be computed using only the values stored in an object. That is, the object providing the service (usually referred to as \texttt{this}) needs information beyond that which it stores. For instance, consider adding a service that computes the distance to a given \texttt{3Dposn} object. In addition to \texttt{this} object another \texttt{3Dposn} object is needed. This other object is unknown when \texttt{this} is constructed and, therefore, cannot be provided as input. In this regard, it is similar to receiving a message given that there is no way to know which messages will actually be received as input. The solution for messages is to make a curried function that consumes the extra input (i.e., a message). The same design tactic may be employed to add services that require extra input. The interface must return a function that consumes the extra input.

To illustrate the technique let us add a service to compute the distance of \texttt{this} \texttt{3Dposn} to a given \texttt{3Dposn}. The first step is to update the interface as follows:
\begin{alltt}
     \textquotesingle{getx}: number
     \textquotesingle{gety}: number
     \textquotesingle{getz}: number
      \textquotesingle{d2o}: number
      \textquotesingle{d2p}: 3Dposn \arrow \ number
\end{alltt}
The data definition of a message is expanded to include \texttt{\textquotesingle{d2p}} for the new distance-computing service. Given that extra input is needed the interface returns a function that consumes the extra input, a \texttt{3Dposn}, and that returns a number for the distance between \texttt{this} and the given \texttt{3Dposn}.

The next step is to refine the \texttt{manager} function to include the new service. This means adding a stanza for the new service to the \texttt{match}-expression as follows:
\begin{alltt}
  ;; message \arrow 3Dposn service throws error
  ;; Purpose: To manage messages for a 3Dposn
  (define (manager m)
    (match m
      [\textquotesingle{getx} x]
      [\textquotesingle{gety} y]
      [\textquotesingle{getz} z]
      [\textquotesingle{d2o}  (dist-origin x y z)]
      [\textquotesingle{d2p}  distance]
      [else
       (error
        (string-append "Unknown message to 3Dposn: " (symbol->string m)))]))
\end{alltt}
The new stanza matches the new message and returns the \texttt{distance} function (yet to be written). The \texttt{distance} function must satisfy the return type specified in the interface definition.

The \texttt{distance} function may now be designed and implemented. Keep in mind that this is a local function inside the \texttt{3Dposn} class. Therefore, this function has in scope all the variables declared in the class. This is where the power of currying is exploited. The previous inputs (\texttt{x}, \texttt{y}, and \texttt{z}) are used to compute the distance to the \texttt{3Dposn} received as input. After following the steps of the design recipe the following local function is added to the \texttt{3Dposn} class:
\begin{alltt}
     ;; 3Dpson \arrow number
     ;; Purpose: Compute distance from this to given 3Dposn
     (define (distance a-3dposn)
       (sqrt (+ (sqr (- x (3Dposn-x a-3dposn)))
                (sqr (- y (3Dposn-y a-3dposn)))
                (sqr (- z (3Dposn-z a-3dposn))))))
\end{alltt}
Observe that this function uses the coordinates of \texttt{this} and of the given \texttt{3Dposn} to compute the distance.

The final step is to develop a wrapper function for the new distance service. As before, this is done following the steps of the design recipe. The resulting function is:
\begin{alltt}
     (define B3DPOSN (make-3Dposn 1 1 1))

     ;; 3Dposn 3Dposn \arrow number
     ;; Purpose: Return distance between given 3Dposns
     (define (3Dposn-distance p1 p2) ((p1 \textquotesingle{d2p}) p2))

     ;; Sample expressions for 3Dposn-distance
     (define ORIGINDP  ((ORIGIN  \textquotesingle{d2p}) ORIGIN))
     (define A3DPOSNDP ((A3DPOSN \textquotesingle{d2p}) B3DPOSN))

     ;; Tests for 3Dposn-distance
     (check-within (3Dposn-distance ORIGIN ORIGIN) ORIGINDP 0.01)
     (check-within (3Dposn-distance A3DPOSN B3DPOSN) A3DPOSNDP 0.01)
     (check-within (3Dposn-distance (make-3Dposn 10 20 30)
                                    (make-3Dposn 2  3  4))
                   32.07
                   0.01)
\end{alltt}
Students are encouraged to first write sample expressions that use the interface because they are new to message passing. They can then perform abstraction over the sample expressions to write the function much like they have done to discover, for example, \texttt{map} and \texttt{filter}.

The reader can appreciate that students are exposed to many prevalent concepts in \oop: interfaces, classes, objects, constructors, observers, and message-passing. The only difference here is that a different syntax is used. Students walk away with an understanding of elements emphasized in \oop.

\section{A Design Recipe for Interfaces}
\label{dr}

Building on the work above, the systematic steps to design the implementation of an interface may now be enumerated. The design recipe for an interface is:
\begin{enumerate}
  \item Identify the values that must be stored and the services that must be provided.
  \item Develop an interface data definition and a data definition for messages.
  \item Develop a function template for the class that consumes the values that must be stored and whose body is a \texttt{local}-expression returning the message-processing function.
  \item Specialize the signature, purpose, class header, and message-processing function.
  \item Design and implement local auxiliary functions needed by the message-passing function.
  \item Write and test a wrapper function for each service.
\end{enumerate}
The first step is problem analysis. It asks to identify the information that specializes each object and the services that an object must provide. For every piece of information identified there must be a parameter. The services must at least include a selector for each piece of information that specializes an object. The second step asks you to develop an interface data definition and a message data definition. There must be a message for each service identified in Step 1.

The third step asks for the development of a function template for the class. Its parameters correspond to the values identified in Step 1 to specialize an object. The name of the class is also the name of the constructor. The function's body is a \texttt{local}-expression that encapsulates all needed functions and that returns the message-processing function. The fourth step has you specialize the definition template for local message-passing function. This function must contain an expression that distinguishes the messages defined in Step 2. The fifth step asks you to develop all the auxiliary functions needed by the message-processing function and make them local. Each auxiliary function is developed using the design recipe.

The sixth step asks for a wrapper function for each service in the interface developed in Step 2. These functions take as input the object that is providing the service and the extra input, if any, required. The body of each wrapper function applies the object providing the service to the appropriate message. If extra input is needed the function returned by the object is applied to the extra input.

\section{Implementing Union Types}
\label{iu}

The implementation of a union type is used to illustrate the steps of the design recipe in practice. We shall use the following data definitions:
\begin{alltt}
     ;; A square, sq, is a an object
     ;;     (make-sq number symbol symbol)
     ;; with a length, a mode, and a color.

     ;; A rectangle, rect, is an object
     ;;     (make-rect number number symbol symbol)
     ;; with a width, length, a mode, and a color.

     ;; A circle, circ, is an object
     ;;     (make-circ number symbol symbol)
     ;; with a radius, a mode, and a color.
\end{alltt}
Based on these definitions a union type for geometric shapes may be defined as follows:
\begin{alltt}
     ;; A geometric shape, gs, is either:
     ;;  1. sq
     ;;  2. rect
     ;;  3. circ
\end{alltt}
Following the steps of the design recipe from \Cref{dr} students are shown how to implement this union type using an interface. A point that is emphasized is that designing interfaces for a union type requires individually reasoning about each subtype. This is not a huge intellectual leap for students given that individually reasoning about each subtype is how functions to process a union type are designed.

\subsection{Step 1: Values and Services}

To implement \texttt{gs} three classes are needed: one for each subtype. The \texttt{sq} class needs to store a number and two symbols. The \texttt{rect} class needs to stores two numbers and two symbols. The \texttt{circ} class needs to store a number and two symbols. This type of analysis is the same as designing an implementation using structures.

The new component is defining the behavior for a union type. The following services are offered by a \texttt{gs}:
\begin{itemize}
  \item Determine if the gs ia a \texttt{sq}, a \texttt{rect}, or a \texttt{circ}
  \item Compute the gs's area
  \item Determine if the area of this gs is larger than a given gs
\end{itemize}

\subsection{Step 2: Interface and Message Definitions}

Based on the services outlined in Step 1, 5 return types and 5 message varieties need to be defined. The interface for a \texttt{gs}, including the messages, is:
\begin{alltt}
     ;; A gs is an interface offering:
     ;;    \textquotesingle{is-sq?}:  Boolean
     ;;  \textquotesingle{is-rect?}:  Boolean
     ;;  \textquotesingle{is-circ?}:  Boolean
     ;;      \textquotesingle{area}:  number
     ;;   \textquotesingle{bigger?}:  gs \arrow Boolean
\end{alltt}
Students easily observe that to compare areas more input is needed and, therefore, a function must be returned.

\subsection{Step 3: Class Function Template}

\begin{figure*}[t!]
\begin{alltt}
     ;; \ldots \arrow gs
     ;; Purpose: Return a gs object
     (define (make-gs \ldots)
       (local
         [   \vdots
          ;; gs \arrow Boolean
          ;; Purpose: Determine if this' area is larger than given gs' area
          (define (is-this-bigger? a-gs) \ldots)

          ;; message \arrow service throws error
          ;; Purpose: Provide service for the given message
          (define (manager m)
            (match m
              [\textquotesingle{is-sq?} \ldots]
              [\textquotesingle{is-rect?}  \ldots]
              [\textquotesingle{is-circ?}   \ldots]
              [\textquotesingle{area}   \ldots]
              [\textquotesingle{bigger?}    \ldots]
              [else
                (error
                  (format "Unknown gs service requested: ~s" m))]))]
        manager))
\end{alltt}
\caption{The Class Function Template for a \texttt{gs}.} \label{classtemp}
\end{figure*}

The class template captures all the similarities that any \texttt{gs} must have. These include the 5 services offered by the interface developed in Step 2. The class function template for a \texttt{gs} is displayed in \Cref{classtemp}. The signature and purpose statement clearly establish that this class returns a \texttt{gs} object. The body of the class is a \texttt{local}-expression that at the very least must encapsulate a function to determine if the area of \texttt{this} object is bigger than the area of a given \texttt{gs} and a manager function to process messages. A function is needed for the \textquotesingle{bigger?} service because more input is required. Experienced students are free to inline a $\lambda$-expression in the body of \texttt{manager} to implement this function if they so desire. The body of the \texttt{local}-expression returns the \texttt{manager} function which is the object that is capable of providing all the services in the interface.

In essence, the template in \Cref{classtemp} is the roadmap students follow to implement the needed classes. In this example, students are explained that it is used to write three classes: one for each \texttt{gs} subtype.

\subsection{Steps 4 and 5: Class Function Template Specialization}

\begin{figure*}[t!]
\begin{alltt}
     (require 2htdp/abstraction)

     ;; number symbol symbol \arrow sq
     ;; Purpose: Return a sq object
     (define (make-sq length mode color)
       (local
         [;; gs \arrow Boolean
          ;; Purpose: Determine if this' area is larger than given gs' area
          (define (is-this-bigger? a-gs) (> (sqr length) (gs-area a-gs)))

          ;; message \arrow service throws error
          ;; Purpose: Provide service for the given message
          (define (manager m)
            (match m
              [\textquotesingle{get-length} length]
              [\textquotesingle{get-mode}   mode]
              [\textquotesingle{get-color}  color]
              [\textquotesingle{is-sq?}     #true]
              [\textquotesingle{is-rect?}   #false]
              [\textquotesingle{is-circ?}   #false]
              [\textquotesingle{area}       (sqr length)]
              [\textquotesingle{bigger?}    is-this-bigger?]
              [else
                (error (format "Unknown gs service requested: ~s" m))]))]
        manager))
\end{alltt}
\caption{The \texttt{sq} Class.} \label{sqclass}
\end{figure*}

Step 4 of the design recipe asks students to specialize the signature, purpose, class header, and message-processing function of the class function template. Step 5 asks students to write the auxiliary functions needed by the \texttt{manager} function. The results of these steps for \texttt{sq} are displayed in \Cref{sqclass}. The class' signature reflects the three types of arguments needed to build a \texttt{sq}. The purpose statement is specialized to explicitly state that a \texttt{sq} is returned. Observe that this is consistent with the overall design because \texttt{sq} is a subtype of \texttt{gs}. The class header defines the constructor's name, \texttt{make-sq}, and has three parameters as suggested by the signature. The specialization of the \texttt{manager} function is the most detailed part of Step 4. Its \texttt{match}-expression processes messages for the components of a \texttt{sq}. Each of these only needs to return the requested value. The rest of the stanzas implement the \texttt{gs} interface. The stanza for \texttt{\textquotesingle{area}} returns the area of \texttt{this} square by squaring its length. The stanza for \texttt{\textquotesingle{bigger?}} returns a function because more input is required. The auxiliary function is named \texttt{is-this-bigger?}. This completes Step 4 of the design recipe.

For Step 5 of the design recipe only one auxiliary function (i.e., \texttt{is-this-bigger?}) needs to be written. Students design this function following the steps of the design recipe for function development. The most salient feature for the purposes of this article is that the wrapper function to compute the area of a \texttt{gs}, \texttt{gs-area}, is used to compute the area of the given \texttt{gs}. This function, of course, is developed in the next step of the design recipe for interfaces. Students are discouraged to follow their first instinct that tells them to use a conditional to implement this service.

The development of the \texttt{rect} and \texttt{circ} classes follows in the same manner. In the interest of brevity their development is omitted. Once all three classes for geometric shapes are implemented students are explained about dynamic dispatch. They are able to see how a service, like \texttt{area}, is always correctly provided because each class locally defines its implementation.

\subsection{Step 6: Wrapper Functions and Tests}

\begin{figure*}[t!]
\begin{alltt}
     (define SQR1  (make-sq   5 \textquotesingle{outline} \textquotesingle{green}))
     (define RECT1 (make-rect 2 3 \textquotesingle{solid} \textquotesingle{blue}))
     (define CIRC1 (make-circ 7 \textquotesingle{outline} \textquotesingle{red}))

     ;; gs \arrow Boolean
     ;; Purpose: Determine if given gs is a sq
     (define (gs-sq? a-gs) (a-gs \textquotesingle{is-sq?}))

     ;; gs \arrow Boolean
     ;; Purpose: Determine if given gs is a rect
     (define (gs-rect? a-gs) (a-gs \texttt{\textquotesingle{is-rect?}}))

     ;; gs \arrow Boolean
     ;; Purpose: Determine if given gs is a circ
     (define (gs-circ? a-gs) (a-gs \texttt{\textquotesingle{is-circ?}}))

     ;; gs \arrow Boolean
     ;; Purpose: Compute area of given gs
     (define (gs-area a-gs) (a-gs \texttt{\textquotesingle{area}}))

     ;; gs gs \arrow Boolean
     ;; Purpose: Determine if first given gs is bigger than second given gs
     (define (gs-bigger? this that) ((this \texttt{\textquotesingle{bigger?}}) that))

     ;; Tests for gs interface
     (check-expect (gs-sq? SQR1)    #true)
     (check-expect (gs-sq? RECT1)   #false)
     (check-expect (gs-sq? CIRC1)   #false)
     (check-expect (gs-rect? SQR1)  #false)
     (check-expect (gs-rect? RECT1) #true)
     (check-expect (gs-rect? CIRC1) #false)
     (check-expect (gs-circ? SQR1)  #false)
     (check-expect (gs-circ? RECT1) #false)
     (check-expect (gs-circ? CIRC1) #true)
     (check-expect (gs-area SQR1)   25)
     (check-expect (gs-area RECT1)  6)
     (check-within (gs-area CIRC1)  153.93 0.01)
     (check-expect (gs-bigger? SQR1 RECT1)  #true)
     (check-expect (gs-bigger? RECT1 CIRC1) #false)
     (check-expect (gs-bigger? CIRC1 SQR1)  #true)
\end{alltt}
\caption{Wrapper Functions and Tests for \texttt{gs}.} \label{wrappers}
\end{figure*}

The wrapper functions are designed following the steps of the design recipe for functions. Students must remember that the input to each wrapper function must contain at least one object (in our case a \texttt{gs} object). The result of this step is displayed in \Cref{wrappers}. In the body of a wrapper function an object must be applied to the appropriate message. If extra input is required then the function returned by the object is applied to said input. This is the case for \texttt{gs-bigger?}. Students also develop wrapper functions to access the fields of each \texttt{gs}-subtype. Observe that this is exactly what needs to be done when fields are private when developing a class, for instance, in \texttt{Java}.

Students are encouraged to define sample instances of each \texttt{gs-subtype}. These are then used to write the tests for the \texttt{gs} interface. This is the strategy followed for the tests in \Cref{wrappers}. Testing must be thorough and all tests must pass. If any tests fails or testing is lacking students must refine their design.

\section{Student Feedback}
\label{fb}

\pgfplotstableread[row sep=\\,col sep=&]{
cat    & prop \\
1 & 0.06 \\
2 & 0.19 \\
3 & 0.19 \\
4 & 0.31 \\
5 & 0.25 \\
}\efftrans

\begin{figure*}[t]
\centering
\begin{tikzpicture}
    \begin{axis}[
            ybar,
            symbolic x coords={1,2,3,4,5},
            xtick=data,
        ]
        \addplot[fill=black] table[x=cat,y=prop]{\efftrans};
    \end{axis}
\end{tikzpicture}
\caption{Distribution of Answers for Effective Transition.}
\label{effective}
\end{figure*}
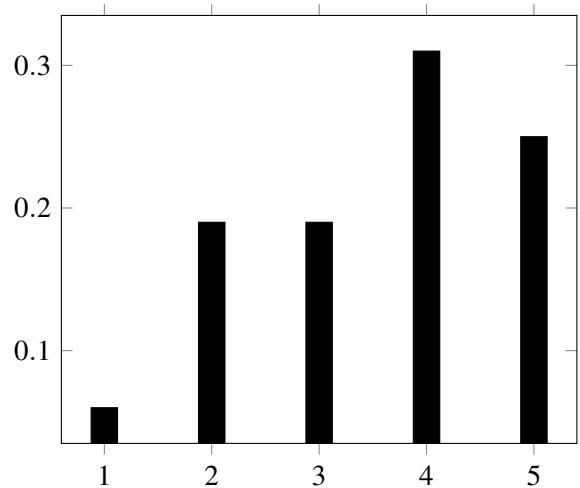

This section presents the first data obtained from students regarding the transition from an \htdp-based course to a \texttt{Java}-based \oop \ course. Students that took the \htdp-based course in the Fall of 2018 were followed-up on after they took their first \oop \ course in the Fall of 2019 or Fall of 2020. A total of 16 students made such a transition and were asked:
\begin{quote}
\emph{How effective do you feel it is to start with How to Design Programs and then move to learning about how to design classes?}
\end{quote}
Students answered on a scale from 1 (not at all effective) to 5 (extremely effective).

\Cref{effective} displays the distribution of responses. We can observe that the distribution is fairly normal (mean = 3.5 and median = mode = 4) with a slight majority of students giving a response above the mean. Overall, students feel that starting with \htdp \ and then moving to \oop \ is effective. 75\% of respondents expressed the approach was effective (responses 3-5). More than half, 56\% of the students felt strongly about the effectiveness of the approach (responses 4-5). We can also observe that the distribution is fairly homogeneous with an IQR = 1.5 (Q1 = 2.75 and Q3 = 4.25). This indicates that there is little variation among the respondents in relation to the effectiveness of the approach.

Qualitatively, students expressed appreciation for being exposed to classes and objects before \emph{learning \texttt{Java}} (their nomenclature). Students that arrived with prior \oop \ experience from high school did express feeling frustrated with the approach at the beginning, but now feel they understand \oop \ much better and are appreciative of their experience.

\section{Conclusions and Future Work}
\label{fw}

This article presents a methodology to introduce beginning students to \oop \ concepts in an \htdp-based course before enrolling in their first \oop \ course. At the heart of the idea is to expose beginning students to object-oriented abstractions using the design concepts and syntax that they have learned. To this end, students are introduced to the idea that a data definition and the valid operations on the defined data are called an interface, that the implementation of an interface is a class, and that an instance of a class is an object. In this manner, specification is separated from implementation. A class is implemented using a message-passing function. The use of functions is natural for students following a program by design curriculum and is intended to foreshadow that classes are functions (as exemplified by \lexps \ in \texttt{Java}). Students learn to implement union types as the process of implementing a class for each subtype. This leads to their initial exposure to polymorphic dispatch. Initial data collected from students suggests that they find the transition from an \htdp-course to \oop \ effective.

Future work includes finding how to make the material more fun for students. Specifically, the implementation of a small video game or simulation would be ideal to spike student interest. In addition, we are exploring how to expose students to abstract classes and inheritance.

\bibliographystyle{eptcs}
\bibliography{Intro-Objects-FP}
\end{document}